\documentclass[11pt]{article}
\usepackage{fancyhdr}
\fancypagestyle{plain}{
  \fancyhf{}
  \fancyfoot[C]{\thepage}
  
}
\usepackage[numbers,sort&compress]{natbib}
\usepackage{amsmath,amssymb}
\usepackage{booktabs}
\usepackage{listings}
\usepackage{xcolor}
\usepackage{adjustbox}
\lstdefinestyle{py}{
    language=Python,
    basicstyle=\ttfamily\small,
    keywordstyle=\color{blue!70!black}\bfseries,
    commentstyle=\color{gray}\itshape,
    stringstyle=\color{purple!70!black},
    showstringspaces=false,
    tabsize=4,
    breaklines=true,
    frame=single,
    numberstyle=\tiny\color{gray},
    captionpos=b
}
\usepackage{hyperref}
\usepackage{cleveref}
\usepackage{needspace}
\usepackage{placeins}
\usepackage{float}
\usepackage{multirow, multicol}
\usepackage{algorithm}
\usepackage[noend]{algpseudocode}
\providecommand{\keywords}[1]{\textbf{Keywords:} #1}

\title{A 60-Addition, Rank-23 Scheme for Exact 3$\times$3 Matrix Multiplication}
\author{%
  Joshua Stapleton\\[-3pt]
  Department of Mathematics, Imperial College London\\
  \texttt{jbs123@ic.ac.uk}}
\begin{document}
\maketitle
\begin{abstract}
We reduce the additive cost of general (non-commutative) $3\times3$ matrix multiplication from the previous records of 61 (Schwartz-Vaknin, 2023) \citep{schwartz2023} and 62 (Mårtensson–Wagner, 2025) \citep{martensson2024} to $60$ without a change of basis. To our knowledge, this represents a new state-of-the-art.
\end{abstract}

\keywords{matrix multiplication; additive complexity; fast algorithms;
ternary weights; verification}

\section{Introduction}
Matrix multiplication runs the modern world. For every word that ChatGPT writes, we estimate that $\approx10^{10}$ small matrix products must be computed.\footnote{Using the token-level FLOP model
$C_{\text{forward}} = 2N$ derived by Kaplan
\textit{et al.}\,\cite{kaplan2020scaling}, a 175 B-parameter model
needs $\approx 3.5\times10^{11}$ fused-multiply-adds per generated
token, i.e.\ $O(10^{10})$ micro-matrix products when mapped to the
32 $\times$ 32 kernels used in modern GPU/TPU GEMMs.} Due to the extreme volume of matrix multiplication required for most digital workflows, even marginal improvements in fundamental matrix multiplication algorithms can translate to substantial cost and energy savings globally. 

In a fast-matrix-multiplication algorithm, multiplicative complexity (often called the rank) is the number of scalar products performed, additive complexity is the total scalar additions or subtractions used to form those products and assemble the result, and arithmetic complexity is simply the sum of the two, i.e.\ the total scalar operations executed. The quest for ever-smaller arithmetic circuits for matrix multiplication has motivated both theoretical bounds and practical kernels since Strassen’s landmark $2\times2$ decomposition. In 1969, Strassen \citep{strassen1969} found a new algorithm that reduced the 'rank' (loosely, the number of multiplications) required for computing an arbitrary 2$\times$2 matrix product from the naive $O(n^3)$ = 8 to 7, initiating an intensive effort to discover faster matrix multiplication algorithms that continues today. Laderman (1976) \cite{laderman1976}, discovered a rank-$23$ algorithm for the $3\times3$ case: the best known upper bound on multiplicative complexity for this problem to this day. With the multiplicative rank seemingly frozen at 23 for almost five decades, later work pivoted to trimming the additive cost of Laderman-type schemes, using graph-theoretic analyses and automated search heuristics \citep{blaser2003,heule2019,martensson2024}. Schwartz and Vaknin subsequently reduced a rank-23 scheme to 61 additions by leveraging an alternative basis and a pebbling heuristic \citep{schwartz2023}.

In recent years, deep-learning-aided algorithm discovery has entered the spotlight. Elser (2016) \cite{elser2016} used a conservative-learning network to 'rediscover' Strassen and Laderman-esque rank-7 and rank-23 schemes respectively. In 2022, DeepMind’s reinforcement-learning system AlphaTensor \cite{alphatensorNature} unearthed new state-of-the-art (SOTA) schemes for the 4$\times$4 case, cutting the multiplicative rank from 49 (Strassen recursion) to 47 scalar multiplications (in $\mathbb Z_2$). Recent results (also from DeepMind) \cite{alphaevolve2025} cut the multiplicative rank for 4$\times$4 complex-valued matrix multiplication from Strassen’s 49 to 48 scalar multiplications, establishing the new upper bound.

In a similar spirit, using basic deep learning techniques, we design a system that sets an upper bound on arithmetic complexity in the $3\times3$ case without a change of basis / altering the algorithm's multiplicative core. More excitingly, this approach is scalable: it can be parallelised, and applied to other algorithms that can be formulated as bilinear mappings.

\section{Neural Discovery Pipeline Overview}
The reduction is achieved by automating the algorithm-generation process through a neural-discovery pipeline. Inspired by Elser, we parallelize the training of small, fully-connected neural networks with architectures reflecting the inherent structure of the algorithm in question. In particular, we force the networks to implement the lowest possible number of arithmetic operations by reducing the dimension of the single hidden layer and encouraging sparsity. Since we can generate an infinite number of random matrix pairs (and their exact products as computed by standard libraries like NumPy) the standard bottleneck of modern deep learning - high-quality data - is not a problem.

At train-time, the bilinear networks are initially unconstrained, allowing the weights to explore a dense floating-point search space and reduce a simple mean-squared-error loss until the underlying physics of matrix multiplication has been grokked. We train the network by standard backpropagation: for each minibatch $(A,B)$ we compute $\hat C=f_\theta(A,B)$ and minimize $\ell(\theta)=\|\hat C-AB\|_F^2+\lambda_T\,\mathrm{SmoothMin}_{\mathcal A}(\theta;T)$ with AdamW~\cite{loshchilov2017}, updating $\theta\leftarrow\mathrm{AdamW}(\theta,\nabla_\theta\ell)$.

Once appropriate accuracy is reached, we activate a 'ternarisation schedule': a temperature-controlled regularizer over a custom loss function that gradually snaps every weight to the ternary alphabet \{-1,0,1\} while preserving accuracy as much as possible; the temperature $T$ is annealed over $R$ steps to push $\theta$ toward $\{-1,0,1\}$. Once the network is $\geq$99\% ternary, we hard-quantize, read out the network's integer-valued weight matrices, and verify symbolically that they implement a valid low-rank matrix multiplication scheme. Finally, a Greedy-Potential post-processor from M\aa rtensson-Wagner reorders the linear parts to minimise redundancies, yielding an exact algorithm. 

Our result tightens the best known arithmetic upper bound for exact $3\times3$ multiplication from $23+61=84$ (Schwartz-Vaknin use a change of basis to achieve this) to $23+60=83$ scalar operations. Unlike Schwartz-Vaknin, our result does not require a change of basis. Additive optimality in the $3\times3$ remains open: we are not aware of any nontrivial lower bound pinning the minimum additions for rank-23. Hence, our result is best described as a new constructive upper bound on the arithmetic cost. 

Pseudocode, implementation, and verification scripts are provided below, and are fully reproducible on commodity hardware. We understand the methodology is not necessarily intuitive, so we also provide a simple animation showing a single forward pass of a matrix pair through a network in the appendix.

\section{Greedy-Potential Reduction Overview}
M\aa rtensson–Wagner introduced a post-processing heuristic, Greedy-Potential, designed to drive down the count of linear operations in an already fixed-rank fast matrix-multiplication scheme without touching its multiplicative core. Starting from the three ternary weight matrices $W_A,W_B,W_C$, the method views each matrix as a signed incidence table $G\in\{-1,0,1\}^{q\times p}$ and repeatedly merges column pairs whenever that substitution reduces the additive cost more than it harms the future saving potential. Their method reached 62 additions without a  change of basis, whereas Schwartz–Vaknin’s alternative-basis variant achieved 61).

Let $G\in\{-1,0,1\}^{q\times p}$ encode $W_A,W_B$ or $W_C$. The potential $P(G)$ counts all reuse patterns $(i,j,\sigma)$ that could be exploited to merge two columns at the cost of a single subtraction. At each step, we choose the triple maximising $M(G,(i,j,\sigma))-1+\alpha P(G')$, where $\alpha=1/10$ worked robustly across all seeds that succeeded in ternarising. Running the routine on the ternarised integer-weight algorithms gave the $60$-add result listed below.

\section{End-to-end neural search and Greedy-Potential reduction pipeline pseudocode}

\begin{algorithm}[H]
\caption{End-to-end neural search and Greedy-Potential reduction pipeline}
\label{alg:pipeline}
\begin{algorithmic}[1]

\Require matrix size $n\!=\!3$,\; rank $r\!=\!23$,\; epochs
         $(E_1,E_2)$,\; loss target $\varepsilon\!=\!10^{-6}$,\;
         ternary ramp length $R$,\; alphabet $\mathcal{A}=\{-1,0,1\}$
\Statex

\State \textbf{Phase 1 (initial exploration)}
\State $\theta \gets \textsc{RandomInit}(n,r)$
\For{$e \gets 1 \;\textbf{to}\; E_1$}                     \Comment{unconstrained}
    \State $(A,B) \gets \textsc{SampleUniform}(n)$
    \State $\hat C \gets \textsc{Forward}(\theta,A,B)$
    \State $\ell \gets \| \hat C - AB \|_F^2$
    \State $\theta \gets \textsc{AdamWStep}(\theta,\nabla\ell)$ \cite{loshchilov2017}
    \If{$\ell < \varepsilon$} 
        \State \textbf{break} \Comment{ready for ternary phase}
    \EndIf
\EndFor

\Statex
\State \textbf{Phase 2 (ternary schedule)}
\For{$e \gets 1 \;\textbf{to}\; E_2$}
    \State $\lambda_T \gets \lambda_{\max}\,\min\!\bigl(1,\tfrac{e}{R}\bigr)$
    \State $T \gets T_{\text{start}} - \bigl(T_{\text{start}}-T_{\text{end}}\bigr)\frac{e}{R}$
    \State $(A,B) \gets \textsc{SampleUniform}(n)$
    \State $\hat C \gets \textsc{Forward}(\theta,A,B)$
    \State $\ell \gets \| \hat C - AB \|_F^2
                 + \lambda_T \,\textsc{SmoothMin}_{\mathcal{A}}(\theta;T)$
    \State $\theta \gets \textsc{AdamWStep}(\theta,\nabla\ell)$
    \If{$\textsc{Ternarity}(\theta) > 0.99 \;\land\; \ell < \varepsilon$}
         \State \textbf{break} \Comment{snap is safe}
    \EndIf
\EndFor
\Statex
\State \textbf{Phase 3 (hard snap \& verification)}
\State $(W_A,W_B,W_C) \gets \textsc{SnapToAlphabet}(\theta,\mathcal{A})$
\State \textbf{assert} $\textsc{VerifyExact}(W_A,W_B,W_C)$
\State \textsc{Save}~$(W_A,W_B,W_C)$
\Statex
\State \textbf{Phase 4 (Greedy-Potential post-processing)}
\State $\textit{scheme} \gets \textsc{fmm.Load}(W_A,W_B,W_C)$
\State $\textit{scheme} \gets \textsc{fmm.GreedyPotential}(\textit{scheme},\alpha=0.1)$
\Comment{23 mults, 60 adds}
\end{algorithmic}
\end{algorithm}

\clearpage

\FloatBarrier                          

\section{A 60-Addition Scheme}
\label{sec:latex-algo}

\begin{figure}[H]                      
  \centering
  \begin{adjustbox}{max width=\textwidth,
                    max totalheight=.90\textheight,   
                    center}
    \begin{minipage}{\linewidth}
\footnotesize      

\[
A=\begin{bmatrix}
A_{0}&A_{1}&A_{2}\\
A_{3}&A_{4}&A_{5}\\
A_{6}&A_{7}&A_{8}
\end{bmatrix}\qquad
B=\begin{bmatrix}
B_{0}&B_{1}&B_{2}\\
B_{3}&B_{4}&B_{5}\\
B_{6}&B_{7}&B_{8}
\end{bmatrix}
\]

\begin{multicols}{2}

\begin{align*}
t_{0} &= A_{0}-A_{3}\\
t_{1}&=A_{4}+A_{5}\\
t_{2}&=A_{6}+A_{8}\\
t_{3}&=A_{1}+A_{2}\\
t_{4}&=A_{7}-t_{1}\\
t_{5}&=t_{0}+t_{2}\\
u_{0}&=B_{0}-B_{2}\\
u_{1}&=B_{4}-B_{7}\\
u_{2}&=B_{1}+u_{0}\\
u_{3}&=B_{5}-B_{8}\\
u_{4}&=B_{6}+u_{3}\\
u_{5}&=u_{1}+u_{2}\\
v_{0}&=M_{4}-M_{14}\\
v_{1}&=M_{2}+M_{22}\\
v_{2}&=M_{7}+M_{21}\\
v_{3}&=M_{9}-v_{0}\\
v_{4}&=M_{10}-M_{18}\\
v_{5}&=M_{3}-v_{1}\\
v_{6}&=M_{5}-v_{2}\\
v_{7}&=M_{12}+v_{3}\\
v_{8}&=v_{4}+v_{7}
\end{align*}

\columnbreak

\begin{align*}
M_{0} &=& -t_{3}\times{}-B_{7}\\
M_{1} &=& (-A_{3}+A_{4}-A_{7})\times{}-u_{1}\\
M_{2} &=& (A_{1}-A_{3})\times{}u_{5}\\
M_{3} &=& -t_{0}\times{}-u_{0}\\
M_{4} &=& -A_{5}\times{}u_{3}\\
M_{5} &=& (A_{8}+t_{4})\times{}B_{7}\\
M_{6} &=& -A_{8}\times{}(-B_{2}+B_{7}+B_{8})\\
M_{7} &=& t_{4}\times{}(B_{5}+B_{7})\\
M_{8} &=& -A_{7}\times{}-B_{3}\\
M_{9} &=& (A_{1}+A_{5})\times{}-u_{4}\\
M_{10} &=& -t_{5}\times{}(B_{2}-B_{6})\\
M_{11} &=& -A_{6}\times{}B_{1}\\
M_{12} &=& (A_{2}-A_{5}+t_{5})\times{}-B_{6}\\
M_{13} &=& (-A_{0}+A_{1})\times{}u_{2}\\
M_{14} &=& -A_{3}\times{}B_{2}\\
M_{15} &=& (A_{6}+t_{0})\times{}(B_{0}-B_{6})\\
M_{16} &=& A_{7}\times{}(B_{4}+B_{5})\\
M_{17} &=& t_{3}\times{}(-B_{6}+B_{8})\\
M_{18} &=& -t_{2}\times{}B_{2}\\
M_{19} &=& -A_{1}\times{}(-B_{3}+u_{4}-u_{5})\\
M_{20} &=& (-A_{1}+A_{4})\times{}B_{3}\\
M_{21} &=& -t_{1}\times{}-B_{5}\\
M_{22} &=& A_{3}\times{}(B_{1}+u_{1})
\end{align*}
\end{multicols}

\[
C=\begin{bmatrix}
M_{19}+v_{5}-v_{8} & M_{0}-M_{13}-v_{5} & M_{17}-v_{8}\\
M_{19}+M_{20}-v_{1}-v_{3} & -M_{1}+M_{16}+M_{22}-v_{2} & M_{21}+v_{0}\\
-M_{3}+M_{8}+M_{15}+v_{4} & -M_{11}+M_{16}+v_{6} & -M_{6}-M_{18}-v_{6}
\end{bmatrix}
\]
\end{minipage}
\end{adjustbox}
\end{figure}

\newpage

\section{Discussion}
Trimming two additions from a 3$\times$3 scheme is unlikely to shift real-world performance without complementary optimisations; the $2/62\approx3.2$\% reduction is negligible for production BLAS (Basic Linear Algebra Subprograms) kernels given that extra memory traffic (register shuffles, cache moves, and pointer chasing) tends to overshadow (or even nullify) any arithmetic saving.

Despite this, our finding strengthens two key points:
\begin{itemize}
  \item \textbf{Greedy-Potential is not saturated.}  The potential value of the $60$-addition scheme remains strictly positive, so further eliminations are theoretically available.

  \item \textbf{Frontier progress does not require hyperscale compute.}
        The entire discovery–snap–reduction pipeline completes in seconds (avg: 5.5s ± 0.1s) on a single consumer‐grade CPU (2023 M3 MacBook Air), showing that careful heuristics and lightweight verification can push symbolic upper bounds without access to giga-GPU clusters, large-scale reinforcement learning infrastructure, or particularly innovative human-aided pipeline design.
\end{itemize}

\section{Conclusion}
We presented a minimal note formalising the $60$-addition Laderman variant discovered by the pipeline, pushing the best arithmetic upper bound for exact $3\times3$ multiplication to $83$ scalar operations. All code and proof artifacts fit in a few dozen lines and run on a laptop. A much more detailed proof and the full discovery methodology will be documented in my MSc thesis, to be released in the coming months.

Future work will target additive and multiplicative frontiers using similar methodologies. We will also target larger problems (eg: 4$\times$4) and explore restricted variants of the problem (eg: in GF(2) / the commutative case). Beyond matrix multiplication, we plan to generalise the ternary-snap pipeline to other bilinear operators: for example, convolutions and finite-field products.

\newpage

\appendix
\section*{Appendix}
\addcontentsline{toc}{section}{Appendix}

\section{Python Verification Script}
\label{sec:python-verify}

\begin{lstlisting}[style=py,caption={Python verification script},label={lst:verify}]
import numpy as np

def fast_3x3_rank23(A: np.ndarray, B: np.ndarray) -> np.ndarray:
    """
    One of the schemes found for 3x3 matrix multiplication with a SOTA arithmetic complexity of 83.

    Parameters
    ----------
    A, B : (3, 3) np.ndarray
        Input matrices (row-major).

    Returns
    -------
    C : (3, 3) np.ndarray
        The product A @ B obtained with only 23 scalar multiplies and 60 additions / subtractions.
    """
    # Flatten to the A0...A8, B0...B8 shorthand
    A0,A1,A2,A3,A4,A5,A6,A7,A8 = A.ravel()
    B0,B1,B2,B3,B4,B5,B6,B7,B8 = B.ravel()

    # pre-additions
    t0 = A0 - A3
    t1 = A4 + A5
    t2 = A6 + A8
    t3 = A1 + A2
    t4 = A7 - t1
    t5 = t0 + t2

    u0 = B0 - B2
    u1 = B4 - B7
    u2 = B1 + u0
    u3 = B5 - B8
    u4 = B6 + u3
    u5 = u1 + u2

    # 23 scalar products
    M0  = (-t3) * (-B7)
    M1  = (-A3 + A4 - A7) * (-u1)
    M2  = (A1 - A3) * u5
    M3  = (-t0) * (-u0)
    M4  = (-A5) * u3
    M5  = (A8 + t4) * B7
    M6  = (-A8) * (-B2 + B7 + B8)
    M7  =  t4 * (B5 + B7)
    M8  = (-A7) * (-B3)
    M9  = (A1 + A5) * (-u4)
    M10 = (-t5) * (B2 - B6)
    M11 = (-A6) *  B1
    M12 = (A2 - A5 + t5) * (-B6)
    M13 = (-A0 + A1) *  u2
    M14 = (-A3) *  B2
    M15 = (A6 + t0) * (B0 - B6)
    M16 =  A7 * (B4 + B5)
    M17 =  t3 * (-B6 + B8)
    M18 = (-t2) *  B2
    M19 = (-A1) * (-B3 + u4 - u5)
    M20 = (-A1 + A4) *  B3
    M21 = (-t1) * (-B5)
    M22 =  A3 * (B1 + u1)

    # v-aggregates
    v0 = M4  - M14
    v1 = M2  + M22
    v2 = M7  + M21
    v3 = M9  - v0
    v4 = M10 - M18
    v5 = M3  - v1
    v6 = M5  - v2
    v7 = M12 + v3
    v8 = v4  + v7

    # outputs
    C0 =  M19 + v5 - v8
    C1 =  M0  - M13 - v5
    C2 =  M17 - v8
    C3 =  M19 + M20 - v1 - v3
    C4 = -M1  + M16 + M22 - v2
    C5 =  M21 + v0
    C6 = -M3  + M8  + M15 + v4
    C7 = -M11 + M16 + v6
    C8 = -M6  - M18 - v6

    return np.array([C0, C1, C2,
                     C3, C4, C5,
                     C6, C7, C8]).reshape(3, 3)

# Test: compare against NumPy for a batch of random matrices
if __name__ == "__main__":
    rng = np.random.default_rng(0)
    for _ in range(1_000):
        A = rng.integers(-9, 10, size=(3, 3))
        B = rng.integers(-9, 10, size=(3, 3))
        assert np.array_equal(fast_3x3_rank23(A, B), A @ B)
    print("All 1,000 random tests passed")
\end{lstlisting}

A single forward pass through a network is visualised in an annotated MP4 and linked in the associated Github repository. The network shown ingests two flattened 3$\times$3 matrices, and computes a matrix product $C'$ using 23 multiplications (as restricted by the dimension of the single hidden layer). A standard mean squared error loss is computed against the true $C$, as computed by a standard library like NumPy. The verification script (Listing \ref{lst:verify}) lives in \texttt{verify.py}; the 20s network-pass video is in \texttt{MatrixMultNetwork3x3.mp4}. The verification script and network-pass animation are available \href{https://github.com/Joshua-Stapleton/60_addition_rank_23_scheme}{here}.  

\newpage

\bibliographystyle{ieeetr}

\end{document}